\documentclass[12pt,preprint]{aastex}

\slugcomment{Submitted to the Astrophysical Journal 26 March 2003 \\ Revised 4 August 2003 \\ To Appear 1 December 2003} 

\shorttitle{Lyman Continuum Emission from Galaxies}
\shortauthors{Malkan, Webb and Konopacky}

\begin{document}


\title{An HST Search for Lyman Continuum Emission \\
    From Galaxies at 1.1 $<$ z $<$ 1.4}


\author{Matthew Malkan, Wayne Webb and Quinn Konopacky}
\affil{Astronomy Division, University of California,
    Los Angeles, CA 90095-1562}



\begin{abstract}
If enough of their Lyman limit continuum escapes, 
star-forming galaxies could be significant contributors to the cosmic background of 
ionizing photons.  
To investigate this possibility,
we obtained the first deep imaging in the far ultraviolet of eleven bright blue 
galaxies at intermediate redshift (1.1 $<$ z $<$ 1.4) 
with the STIS FUV MAMA detector on the Hubble Space Telescope. 
{\it No} Lyman continuum emission was detected.   
Sensitive, 
model-independent, upper limits of typically 
$<$ 2$ \times $10$^{-19}$ ergs cm$^{-2}$ sec$^{-1}$ \AA$^{-1}$ 
were obtained for the ionizing flux escaping from these normal galaxies.
This corresponds to lower limits on the observed ratio of 1500 to 700$\AA$ flux of 150 up to 1000.
Based on a wide range of stellar synthesis models, this suggests that 
less than 6\%, down to less than 1\%, of the available ionizing flux emitted by hot stars 
is escaping these galaxies.  The magnitude of this spectral break at the Lyman limit confirms 
that the basic premise of 
``Lyman break'' searches for galaxies at high redshift can also be applied at intermediate 
redshifts.    
This implies that the integrated contribution of galaxies to the 
UV cosmic background at z $\sim$ 1.2 is 
less than 15\%, and may be less than 2\%. 
   
\end{abstract}


\keywords{ultraviolet: galaxies -- intergalactic medium -- diffuse radiation}


\section{THE DIFFUSE IONIZING BACKGROUND}

Blueward of the Lyman limit lies a wavelength region of
great observational uncertainty and 
astrophysical significance.  
These far-UV photons control ionization of  
the interstellar and intergalactic 
media today, and at all times since recombination \citep{mad99}. 

The origin of this UV radiation field 
that re-ionized the early Universe was considered by 
\citet{sar80}. 
Quasars are known to be strong contributors to the cosmic UV background, as 
their spectra 
show little intrinsic fall-off across the Lyman limit (e.g., \citet{sun89}). 
\citet{bec87} 
calculated the diffuse 
ultraviolet radiation field using assumptions about the quasar luminosity function 
and the shape of the intrinsic UV spectrum of quasars.  They also discussed the 
possibility that young galaxies could be significant contributors 
to the UV background. 

In addition to measuring the Lyman continuum from sources directly, 
the strength of the UV background may be inferred 
using the ``proximity effect" -- the trend toward smaller number 
densities of Ly$\alpha$ lines in a quasar absorption spectrum 
as one approaches the redshift of a background quasar. Using this 
method \citet{baj88} 
calculated a larger UV background than the 
integrated emission of known quasars at z $>$ 3, 
requiring ``appreciable sources of ionizing photons other than quasars".  
Since quasars have a lower space density than galaxies,
the ionizing photons from more uniformly distributed 
normal galaxies would set a floor on the {\it fluctuations} in the
ionizing background radiation from one location to another.  

A related method of estimating the UV background is by measurement of the 
Ly$\alpha$ forest decrement as a function of redshift \citep{mad95}. 
\citet{haa96} 
pointed out that re-radiation from intergalactic absorption systems acts as sources, 
as well as sinks, of ionizing photons which affect the amplitude, shape, 
and fluctuations of the UV background.  They find this stochastic reprocessing 
of quasar light appears to provide enough ionizing photons to account 
for the proximity effect at high redshifts.

\subsection{Predicted Escape of Ionizing Photons from Normal Galaxies}

{\it Redward} of the Lyman limit most of the diffuse cosmic UV radiation 
comes from the 
numerous normal galaxies, particularly those with high instantaneous
star formation rates, e.g., \citet{gia97,fre99}. 
{\it Blueward} of the Lyman limit, the amount of ionizing 
photons from hot stars that escape normal galaxies 
is not well known on either theoretical or observational grounds.  
Locally it is poorly measured because of our inability to
detect it through the HI absorption of the Milky Way.
The intrinsic spectra 
of hot stars around the
Lyman limit are also not accurately measured \citep{lan03}.
If even small amounts of radiation blueward of the Lyman limit escape from
normal galaxies
(e.g., $\sim$ 10\% of their mid-UV flux), then (in aggregate) galaxies could still be the 
dominant source of the UV background \citep{set97,deh97} 

Researchers use the term ``escape fraction" in several model-dependent ways. 
The most conceptually straightforward definition
is the number of ionizing photons escaping the 
galaxy divided by the total number of ionizing photons emitted by stars within it,
i.e., f$_{esc}$ = N$_{esc}$/N$_{emit}$. 
However, to avoid any dependence on models, the {\it observed} 
limit of 1500 to 900 \AA\ flux
is sometimes quoted, for comparison with the value which would be expected if all 
ionizing photons escaped.

\citet{cas02} 
used photoionization models of well studied extragalactic giant H II regions and predicted  
escape fractions of ionizing photons 10\% $<$ f$_{esc}$ $<$ 73\%.  
\citet{cia02} 
used 3-D numerical simulations to predict escape fractions  to be as high as 60\% 
from Milky Way-like galaxies with high star formation rates.
In a recent study of the influence of supershells and galactic 
outflows, \citet{fuj02} 
conclude that {\it dwarf} starburst galaxies may 
have played an important role in ionizing the universe at z $>$ 5 with total escape 
fractions f$_{esc}$ $>$ 20\%.  

The simple theoretical picture of a
calm, uniform disk galaxy, fully shrouded in H I and opaque to ionizing photons,
is unlikely to be realistic, particularly in the younger Universe.
In the Milky Way, radio maps reveal large-scale 
bubbles and chimneys that appear to have been blasted through the 
interstellar gas of the disk by the combined effects of many supernovae and 
vigorous stellar winds (Heiles 1987; also in M82 Devine \& Bally 1999). 
These evacuated or ionized holes may open up escape paths for ionizing photons 
heading out of the disk \citep{roz99,kun98}. 
Detection of ionized gas well above the plane of the Milky
Way (the ``Reynolds layer" and the Magellanic Stream) has been interpreted 
as meaning that 5 - 10\% of hydrogen ionizing photons escape from our
own galaxy \citep{rey85,bla99}. 
It is also possible that the detection of strong ionized calcium absorption along most 
lines of sight through the Milky Way halo
requires the escape of a substantial number of ionizing photons 
from the disk \citep{sav88}. 

\subsection{Previous FUV Observations of Galaxies}

HUT obtained strong upper limits on the Lyman continuum escaping from 3 low-luminosity,
low-redshift galaxies
(and less sensitive observations of a fourth galaxy)
\citep{lei95}. 
After correction for Milky way absorption, these observations translate into
five-$\sigma$ flux upper limits of 
F$_{\lambda ~ 900\AA}$ $<$ 4$ \times $10$^{-15}$ ergs sec$^{-1}$ cm$^{-2}$ $\AA^{-1}$
\citep{hur97,fer01}. 
Based on models, the implied fraction 
of escaping Lyman continuum photons f$_{esc}$ $<$ 8 --- 28\% (five-$\sigma$ upper limits).
FUSE has produced a similarly strict flux upper limit \citep{deh01}, 
for Mkn 54 (z = 0.0448) with F$_{\lambda ~ 900\AA}$ $<$ 1$ \times $10$^{-15}$ 
ergs sec$^{-1}$ cm$^{-2}$ $\AA^{-1}$. This can be converted into a five-$\sigma$ upper limit
to the ionizing escape fraction of 10\%.
 
These results at low redshifts may not apply to earlier times when galaxies 
may have had higher UV luminosities and lower UV extinction.  
At higher redshifts, galaxies appear to be bluer and more disturbed 
(e.g., \citet{wil96}) 
suggesting the possibility that more of their Lyman continuum photons
may have escaped at earlier cosmic times.  
Given the 
effects of intergalactic absorption, detecting these photons may become more difficult above
$ z \sim 2$.
With the new generation of 10-meter class ground based telescopes one can 
now observe the rest frame Lyman continuum of galaxies at redshifts $\sim$ 3.
Steidel, Pettini and Adelberger (2001, SPA) 
reported the detection of Lyman continuum flux in the composite 
spectrum of 29 Lyman break galaxies (LBGs) with $<z>$ = 3.4. 
Since the galaxies in their  
sample were drawn from the bluest quartile of LBG spectral energy distributions,
they are likely to be younger and less dusty than their 
contemporaries. Nevertheless, if their measured ratio 
of rest frame flux density at 1500$\AA$ to that in the Lyman 
continuum,  F$_{\nu 1500\AA}$/F$_{\nu 900\AA}$ = 4.6 $\pm$ 1.0 were typical, 
then galaxies at z = 3 
would produce about five times more H-ionizing photons per co-moving volume than 
would quasars.   Given the difficulties of the 
observation and the uncertainties in the intervening medium, 
SPA describe these results as preliminary.  Contradicting this result 
are the observations of \citet{gia02} 
who failed to detect the Lyman continuuum in deep
VLT observations of several of Steidel's brightest Lyman break galaxies.
In addition, \citet{fer03} observed 27 HDF galaxies at 1.9 $<$ z $<$ 3.5, and concluded 
that no more than 15\% of the ionizing photons escaped.
Whether or not one accepts the tentative Steidel 
detection of Lyman continnuum light from the brightest bluest galaxies at 
z $\sim$ 3, that does not settle the question of 
the emission of ionizing radiation from other types of galaxies, at other redshifts. 

\section{THE OBSERVATIONS} 

\subsection{Object Selection}

We searched the literature and the NASA/IPAC Extragalactic Database (NED) 
for all known (as of 1997) bright (B $\le$ 23.0), non-active
galaxies at {\it spectroscopically confirmed} redshifts in the range 1.1 $<$ z $<$ 1.7.  
The lower redshift bound was selected to put all emission redward of 912$\AA$ out of 
our detection passband.  The upper redshift limit was chosen to minimize the 
effects of intergalactic absorption.
We further selected among the candidates based upon
colors, choosing the bluest, brightest, galaxies, since
the sensitivity of our search is directly proportional to the
strength of the flux emitted just longward of the Lyman limit. 
We avoided all targets that might be active galaxies based upon the presence 
of high-ionization emission lines (e.g., other than [O II] at 3727$\AA$).  
Because of the sharply decreasing numbers of spectroscopically confirmed
galaxies as z increases towards 2, 
our brightest candidates were found in the redshift interval 1.1 $<$ z $<$ 1.4.  
Given the limited photometry available for most of these galaxies, they were in effect 
selected based upon their brightness in the shortest observed passband,  
be that U or B. Four CFRS galaxies were selected for their brightness in V, combined with
their extremely blue V-I colors.    

Our sample size of eleven was a compromise between examining a large enough number 
of sources to account for possible line-of-sight (LOS) variations and integrating long 
enough on each object to set very significant flux limits.  
Several different geometric possibilities exist by which normal 
galaxies could be significant cosmic contributors of ionizing
photons:  a) most galaxies emit Lyman continuum isotropically, but at
a low level; b) most of them
emit virtually no Lyman continuum photons {\it except} along a few,
unusually clear lines-of-sight; or c) most of them emit virtually
no Lyman continuum photons whatsoever, but a minority of 
special galaxies (e.g., irregular or disturbed galaxies) do emit.
Deep integrations on only a few objects are required to 
detect the Lyman continuum in hypothesis a), while   
a broad survey of many galaxies, each of which samples one line-of-sight
from Earth, is needed for testing hypotheses b) and c).
Hence, we employed a combination of both ``deep" and multiple ``broad" observations,  
to test each of these possibilities.  
The targets 
we selected, their redshifts, our pointing coordinates, dates of observation, 
foreground 100$\mu$ cirrus brightness, 
and exposure times are given in Table 1.   


\subsection{Description of the Observations}

Observations were conducted with HST between October 2000 and October 2001. 
The targets were imaged with the Hubble Space Telescope (HST) 
using STIS 
with its FUV MAMA detector, under Guest Observer Program 8561.  
The field of view consists of 1024 x 1024 pixels of 0.0247 arcseconds
each, spanning a total of 25 arcseconds. 
The longpass F25SRF2 
filter was chosen to provide high throughput and strict rejection of sky 
background photons.
Transmission is characterized by a sharp turn-on peak just short of 1300$\AA$, 
which tapers 
off steadily until just past 1900$\AA$, with a central wavelength of 1480$\AA$.   
The CsI photocathode will not detect any light redward of the Lyman limit, in the 
target's rest frame,  
because of the instrument's steep cutoff at 
$\lambda \ge 1925\AA$.  Red leak is not a problem at the flux levels we are 
observing.  

In measuring these faint sources it was important to account for and minimize
the sources of background. The count rate from the MAMA dark current was predicted 
in the instrument handbook
(www.stsci.edu/hst/stis/documents/handbooks/currentIHB/)
to be 5 -- 10 x 10$^{-6}$ counts sec$^{-1}$ pixel$^{-1}$. 
The sky background we observed was small, typically 
$\sim$ 0.1 counts per 3000 seconds (somewhat higher than the dark current).
Photon noise dominated the uncertainty in our measurements.  
Comparisons of the mean and standard deviations of counts 
per pixel in large and small 
blank sky regions were fully consistent with Poisson statistics, 
$\sigma$(N) = $\sqrt{N}$ where N is the total number of 
background photons detected in an aperture.  

Earthshine provided the dominant background.  The background from the 
Earth in shadow was   
roughly two orders of magnitude less 
than the average background from the illuminated Earth.
Observations were primarily conducted in the Earth's shadow.  
Ten of the eleven galaxies 
were observed in ``TIME-TAG" mode which records the arrival times of each individual photon.
This information allowed us to optimize signal to noise ratio by 
excluding any data taken during periods of high background brightness 
(e.g., observations in the presence of the illuminated Earth).  
In TIME-TAG, the recorded photons are binned
in high spatial resolution mode, where 2048 x 2048 pixels are each 0.01237 arcseconds wide. 
One galaxy, SSA 22-16, was observed in the ACCUM mode, in which
all the photons over the entire 5598-second night-time portion of the orbits were
included in the final image.  

\section{DATA ANALYSIS PROCEDURES AND PRODUCTS}

\subsection{Data Reduction}

	We identified the time periods of each exposure with the darkest 
backgrounds by first 
averaging the count rates (over the entire field-of-view) in intervals of 
100 seconds.  Typically, the 
period of minimum sky count rates, during the night-time portions of 
each orbit, persisted for about 2000 seconds.  We then summed all the photons measured during
those dark periods, using the inttag task in Version 2.11.3b of IRAF. 
These total ``dark" integration times, listed in Table 1, ranged from 3500 up to 6300 
seconds for different objects.  The average count rates for dark sky background  
ranged from 
0.032 to 0.15 counts second$^{-1}$ pixel$^{-1}$. On average, exclusion of the data 
taken with earthshine background, which was typically 25 - 40\% of the total exposure time, 
improved the flux limit slightly (lowering it by approximately 25\%). 

Sensitivity variations 
across the FOV were corrected by dividing the images by standard flatfield calibration frames. 
As noted in the STIS manual, there are smooth but reproducible variations across the field. 
In our case those peaked at roughly a 15\% increase in sensitivity near the upper left edge 
of the field.  Photometry of possible sources was performed in circular apertures 
of different radii with standard 
sky subtraction.  Counts were converted to fluxes as specified in the STIS instrument handbook 
at the rate of 1 count sec$^{-1}$ corresponding to a flux of 4 x 10$^{-17}$ ergs cm$^{-2}$ sec$^{-1}$.

As viewed at intermediate 
redshifts, the small angular size of galaxies concentrates their light on only a small number of 
FUV MAMA pixels.  Available WFPC2 imagery 
for several of the normal galaxy targets indicates the typical 
FWHM $\sim$ 0.5 arc seconds (Lubin et al. 1998) or 21 FUV MAMA pixels in radius.
To estimate the angular size of {\it compact} galaxies at redshift z $>$ 1, 
we used an archived WFPC2 image
of the SSA22 field.  Measurements of SSA22-16, SSA22-10 
(a compact galaxy), and SSA22-20 (a relatively isolated object in the upper left 
of the image) gave us a 
half radius for compact galaxies of 5 FUV MAMA pixels,
appropriate for marginally resolved objects close in angular size to the PSF of the camera.
In what follows, we quote some results for the smaller angular size compact galaxies but,  
conservatively, our tables and flux limits are specified for normal-size galaxies at 
these redshifts.   

\subsection{Effects of the Intervening Medium on the Intrinsic Spectra}

Unlike at higher-redshifts, the opacity of the intergalactic medium (IGM) is not significant 
at the redshifts we are observing. 
Quasar absorption features can be used to estimate the decrement due to H I absorption 
as a function of wavelength and redshift \citep{wey98,imp96,mad95}.
At z = 1.2 the Ly$\alpha$ line density per unit redshift, normalized for rest EW = 0.24$\AA$, 
is $\sim$ 35.  Integrating over line equivalent width produces EW(integrated) $\sim$ 10$\AA$ 
of absorption at a wavelength of 1200$\AA$.  Thus the resulting continuum flux decrement 
is $\sim$ 1\%, negligible compared to our other sources of error, with Ly$\alpha$ 
being the major opacity source.  We note that the corresponding uncertainty
in correcting for IGM absorption at a redshift of 3 is far larger, and seriously
affects conclusions about escape of ionizing photons from galaxies at that
high a redshift.

Corrections were made for the effects of Milky Way extinction.  Using the IRSKY 
tool (www.ipac.caltech.edu/ipac/services/irsky.html) we computed the IR sky backgrounds for our target fields, 
and applied the scaling E$_{B-V}$ = 0.018 x I$_{100\mu}$(mJy/Sr) \citep{sch98}.  
The reddening for our target fields was then estimated based upon
a standard Seaton Milky Way reddening law \citep{sea79} with 
a ratio of total to selective extinction, A$_V$/E$_{B - V}$ = 3.   

\section{RESULTS}

\subsection{Inspection of the Final Images -- Two Sources}

	Examination of the eleven final images revealed that only two sources were detected -- 
and neither was a target galaxy.  
Both were identified as lower-redshift galaxies in the fields of view of
the higher-redshift galaxies.  The first of these detections was in the ACCUM mode 
exposure targeted at SSA22-16, and appears to be SSA 22-10 at z = 0.132,
located 4.5 arcseconds east (left in our image) of SSA 22-16.  
Figure 1 compares a WFPC2 visible image of SSA 22-16 and SSA 22-10 (right)  
with our STIS ultraviolet exposure of the same field (left).  The images have 
a common plate scale 
(the arrows are 4.5 arcseconds long) and are both oriented with north at top.  
The lower redshift galaxy SSA 22-10 is visible in the UV but our
targeted galaxy SSA 22-16 is {\it not}.
After subtraction of a sky value 
of 0.1241 counts pixel$^{-1}$, the image of SSA 22-10 contained 720 counts within an aperture of radius 
of 65 pixels.  This corresponds to a flux level of 5 x 10$^{-18}$ ergs cm$^{-2}$ sec$^{-1}$ $\AA$$^{-1}$
at $\sim$ 1650$\AA$. 
This is about ten times brighter than its I-band measured flux of 5.4 x 10$^{-19}$, 
making SSA 22-10 a bluish galaxy with F$_{\lambda}$ $\sim$ $\lambda$$^{-1.4}$.  

	A second source was detected in the image targeted at HDF-IW4-1002.1353. 
The source appeared in the lower left corner of the frame, and 
is believed to be the optically faint, X-ray source J123627.2 +621308, 
and is located 0.3 arcseconds east and 12.8 arcseconds north of the 
target.  The net counts from this object were 
determined to be 293 at a radius of 50 pixels and a sky value of 0.15.  This corresponds to 
3 x 10$^{-18}$ ergs cm$^{-2}$ sec$^{-1}$ $\AA$$^{-1}$, or about 30 times its I-band flux of 10$^{-19}$.  This suggests 
an inverse wavelength squared dependence in F$_{\lambda}$, or a flat spectrum in F$_{\nu}$, 
probably from a quasar.  

We compared the radial growth of the number of counts in the two detected sources 
with each other and with WFPC2 images we had on hand of our targeted 
intermediate-redshift galaxies.  We used these observed profiles to estimate how the FUV light 
from our target galaxies should have been distributed spatially to better estimate 
our upper limits.  The two sources had quite 
similar curves of growth, with half of the counts contained within a radius 
of $\sim$ 20 pixels.  Not unexpectedly, the light is less concentrated in the centers of 
our detected sources than it is in the WFPC2 images of our targeted intermediate-redshift galaxies.

\subsection{New Lyman Continuum Flux Upper Limits for Galaxies}

No Lyman continuum light was detected from any of our eleven targeted sources, setting new 
flux upper limits on the potential contributions to the UV background from normal galaxies 
at 1.1 $<$ z $<$ 1.4.  The first column of Table 2 contains our dereddened (first row) and 
observed (second row) upper bounds on the Lyman continuum flux for 
normal-size galaxies at these intermediate redshifts. 
Sizing the apertures accordingly (the flux limit scales linearly with the aperture diameter), 
those flux upper limits are $<$ .005 - .015 $\times $10$^{-30}$ ergs cm$^{-2}$ sec$^{-1}$ Hz$^{-1}$ 
for compact galaxies and $<$ 
.03 - .10 $\times $10$^{-30}$ ergs cm$^{-2}$ sec$^{-1}$ Hz$^{-1}$ for 
normal-size galaxies.  These findings apply to the observed flux levels, and are independent of model assumptions 
that predict the actual fraction of Lyman limit photons that escape from normal galaxies.  

\subsection{Co-Adding the Images to Lower the Detection Threshold}

If one assumes that all the candidate sources are similar and that they are emitting isotropically, 
one can combine our measured upper limits into a single statistical result for these   
intermediate-redshift blue galaxies.
For this test, all 10 TIME-TAG images were added together to search 
for weak flux from the composite of all target measurements. 
Information on pointing error for STIS was not available, but was determined for WFPC2 to be 
0.86 arc seconds. 
This error, combined with the uncertainty in the positional coordinates, means 
one can only coregister these images to on the order of a typical normal galaxy diameter 
at these redshifts.  
In fact, a couple of marginally significant signals 
were detected in the co-added image, but these were {\it not} at the center of  
the FOV where the galaxies were targeted.  They were off-center and appear to be low-level 
blemishes in the STIS camera. One of the apparent peaks in counts, centered around pixel [476,718], 
is present in the same location in about half of the TIME-TAG images (of different sources), and is therefore 
clearly not a cosmic source.  Closer to the center of the FOV, where the targets were centered, at pixel [506,576] 
is another blemish.  It appears at the same pixel location on several different target frames, and 
subsequent photometry showed the signal, even in a generously sized aperture, to be less than 
a three-sigma departure from the sky level.  
{\it Assuming} adequate co-registration of these 10 frames, 
the upper limit on FUV flux from the targets in this
composite image, with total integration time of 42,400 seconds, 
is about 2.3 times as sensitive as in individual fields and produces a flux upper limit 
$<$ 10$^{-20}$ ergs cm$^{-2}$ sec$^{-1}$ $\AA$$^{-1}$.  

\subsection{Observed Spectral Energy Distributions - The F$_{\lambda < 912\AA}$ / F$_{\lambda > 912\AA}$ Lyman Break}


Internal to the galaxy, O and B stars produce a flux of Lyman continuum 
photons, only a small fraction of which are likely to escape into the IGM.
At the edge of the galaxy, we refer to the flux immediately shortward 
of the Lyman continuum, as F$_{o,~\lambda < 912\AA}$, 
and immediately longward of the Lyman continuum, as F$_{o,~\lambda > 912\AA}$. 
In the spectra of galaxies, the discontinuity or ``break" between these values 
is caused by the enormous optical depth of neutral hydrogen in the interstellar medium.  
Fluxes on both sides of the Lyman continuum limit are reddened to the observer, by dust internal 
to the source galaxy, our Milky Way, and by passage through the IGM, resulting in {\it observed fluxes}, 
immediately above and below the Lyman continuum, of F$_{\lambda < 912\AA}$ and F$_{\lambda > 912\AA}$.  
Published UBVRI-band photometry of the galaxies observed here establishes their baseline 
continuum strength and shape (in the rest frame mid-UV).  Photometric errors are not quoted 
in the literature for these objects \citep{tam01,lil91,cow96}.
We have assumed 10\% photometric errors, and made the height of the symbols in Figure 2
equal to this range of uncertainty. 
The strength of the dereddened 
F$_{o,~\lambda < 912\AA}$/F$_{o,~\lambda > 912\AA}$ 
break is shown in the spectral energy distributions (SEDs) 
of these objects (Figure 2).  

The actual escape fraction of 
photons that leak out of a galaxy is dependent upon, among other not well-determined things, 
the spectral shapes and internal reddenings of the sources. 
One previous formulation (e.g., SPA) is in terms of 
observed quantities as a {\it relative} escape fraction, 
f$_{rel~esc}$ $=$ 3. x F$_{\nu, 1500\AA}$/F$_{\nu, 900\AA}$.  
One then uses observations of the F$_{\nu, 1500\AA}$/F$_{\nu, 900\AA}$ ratio, in a model-independent way, 
to specify the relative escape fraction.  At high redshift, 
one should also include any flux decrement from 1500 to 900$\AA$ due to the IGM.  
The factor of 3 arises in the definition from the anticipated {\it intrinsic} drop in 
flux from 1500$\AA$ to 900$\AA$, which depends upon assumptions about 
the IMF and stellar ages.  We investigated this dependence over a range of 
Bruzual-Charlot stellar synthesis models \citep{bc93}, 
in which the galaxy originally formed at z $\sim$ 10.  The simulations included cases with 
both older (z $\sim$ 1.8), and  
more recent but smaller, starbursts (z $\sim$ 1.4 - 1.7). These starbursts, of duration one gigayear, 
produced all the O and B stars.  
Figure 2 includes two of these models, one blue and red, at two flux levels which 
bracket most of our data.  They provide a 
reasonable match to the 
published photometry for most of our target galaxies, and demonstrate the insensitivity of  
the Lyman continuum levels to model choices.
Two extreme stellar population models were generated,
using the Bruzual Charlot 1995 code.
In both models, the original stellar population is
assumed to have formed early, at a redshift of 10.
Then a second burst of star formation starts at z=1.6.
Both bursts are described by the BC95 1 Gyr constant
star formation rate model, with an assumed Salpeter IMF.
No internal reddening has been added.
Adding any reddening to the model would have resulted in
an even younger intrinsic stellar population, which would
have relatively stronger Lyman limit emission than in the models in
Figure 2.

The galaxy model is observed at our typical redshift of 1.3.
For the H=75 open cosmology assumed, this means that the
starburst has been going on for about 0.6 Gyr at the time
of the observation. Models without ongoing star formation
do not reproduce the relatively strong, flat UV
continua of these luminous blue galaxies. Aside from arbitrary vertical
luminosity normalizations, the two models we plot differ
in the fraction of the galaxy mass incorporated into the
second, more recent star formation episode.
The redder model has 2\% of the total galaxy
mass in the z=1.6 starburst.  The bluer model
(heavier line) has
equal mass in the young and the old stellar components.
For clarity, two plots of each model are shown, separated
by 0.4 dex.  This approximately brackets the range of
brightnesses and colors of our target galaxies.
These two models do not provide a complete search of
all the possible range of parameter space for stellar
synthesis models.  Nonetheless, they cover wide enough
extremes to give a good idea of what the far-UV slope
in our program galaxies is likely to be.
Note that in spite of enormous differences between the
weak- and strong-burst models at visible wavelengths,
they are very similar in the far-UV.  This is because
in both cases the observed far-UV continuum is dominated
by the same population of the hottest, youngest main
sequence stars.   

If one assumes an intrinsic flux density drop F$_{\nu, 1500\AA}$/F$_{\nu, 900\AA}$ $\sim$ 3, 
and given that we have no flux decrement 
from the IGM at these redshifts, our typical observed limits would imply a relative escape fraction 
f$_{rel~esc}$ $\le$ 0.3\%.  While the intrinsic flux density ratio F$_{\nu, 1500\AA}$/F$_{\nu, 900\AA}$ is 
typically a factor of 3, it can range from 2.5 to 5.5 and beyond.  
The first column of Table 2 contains our dereddened measurements of the  
upper bounds on Lyman continuum flux (detection limits) for 
normal size galaxies at these intermediate redshifts. 
To match the observed SED's of our target galaxies, we conservatively chose an 
intrinsic flux density ratio F$_{\nu, 1500\AA}$/F$_{\nu, 900\AA}$ $\sim$ 10, in the middle of 
the range recommended by Leitherer and Heckman (1995),  
resulting in a typical limit of f$_{rel~esc}$ $\le$ 1\%.

\section{DISCUSSION}

\subsection{Validation of the Lyman Break Method}  

The Lyman break technique is the most powerful
photometric method known for identifying
galaxies at high-redshift.  It is based on the absorption of their
Lyman limit emission as the unique identifying signature which
separates them from the {\it much larger} foreground population of galaxies 
\citep{ste92}. 
Especially at the lower redshifts, intergalactic absorption is not the only source of this 
effect.  One also assumes that the {\it intrinsic}
spectrum of the star-forming galaxy drops substantially blueward of its
Lyman limit, due to heavy internal H I absorption \citep{mad95}. 
Even with near zero IGM absorption, our observations confirm that the intrinsic 
Lyman limit drop in the spectra of blue galaxies is at least 3 to 4 magnitudes. 
Even if the Lyman break occupies only a small part of the total filter bandwidth (e.g., 
objects at z $\sim$ 2 observed in the U filter), it will be detectable in typical 
Lyman break photometric color searches. 

\subsection{The Contribution of Galaxies at 1.1 $<$ z $<$ 1.4 to the UV Radiation Field}

We used our observed flux upper limits and the Bruzual Charlot models\footnote{The intrinsic 
spectra of hot stars around the Lyman limit are not well measured.  
The recent EUVE detection of $\beta$ CMa at 500-700$\AA$ \citep{cas96} 
revealed a factor of 
20 times more ionizing luminosity than had been predicted by current stellar atmosphere models (which 
underestimated the importance of non-LTE effects).} to estimate a co-moving 
emissivity, following Madau (1995; 1997).  
In this intermediate redshift range, 
where the cosmic emissivity from galaxies is peaking, one predicts 
an intrinsic emissivity from all galaxies of 6.26 x 10$^{25}$ ergs sec$^{-1}$ 
Hz$^{-1}$ Mpc$^{-3}$ 
at our measurement rest wavelength of approximately 
700$\AA$.  Since less than 1 to 6\% of this luminosity escapes the galaxy at this wavelength, 
we find the co-moving emissivity from galaxies at intermediate redshfits (z $\sim$ 1.2) to be less than 
4 x 10$^{24}$ down to 6 x 10$^{23}$ erg sec$^{-1}$ Hz$^{-1}$ Mpc$^{-3}$, depending on which end 
of the escape limit range applies. 

\citet{haa96} 
quote a volume emissivity from quasars near the H I ionization edge of 
3 x 10$^{25}$ ergs sec$^{-1}$ 
Hz$^{-1}$ Mpc$^{-3}$ at z $\sim$ 1.2.  At intermediate redshifts, 
our new Lyman continuum upper limits suggests 
that galaxies make up less than 15\%, 
and possibly less than 2\%, of the cosmic ultraviolet background.  
While less than 1--6\% of the ionizing flux escaped from these {\it bluest} of galaxies, 
it is possible 
that this upper limit will be even lower for the general population of galaxies at 
1.1 $<$ z $<$ 1.4.

\subsection{Comparison of Our Flux Upper Limits with the Literature} 

	SPA remains the only reported detection of Lyman continuum from galaxies, albeit 
at z $\sim$ 3 - 4, with f$_{rel~esc}$ $\ge$ 50\%.  We do not detect the same effect for intermediate redshift 
galaxies, where we find f$_{rel~esc}$ $\le$ 1\%.  It remains possible that at higher redshifts, higher star formation rates 
occuring on global scales produce interstellar media that may leak ionizing radiation. 
In addition, as was pointed out by \citet{hec01}, 
the Lyman Break Galaxy composite spectrum  
employed by SPA showed signs of outflow in the 400 to 500 km sec$^{-1}$ blueshifted interstellar 
absorption lines.  Such outflows may be a necessary signature of galaxies leaking ionizing radiation, but we 
do not know how strong such a signature is in our target galaxies.  

	\citet{gia02} 
point out that the escape fraction for ionizing photons reported by SPA  
would allow the galaxies {\it alone} to produce a cosmic ultraviolet background flux which exceeds the predictions 
of the proximity effect from the Ly$\alpha$ forests of quasars. {\it If}, however, 
galaxies at redshifts of 2 or 3 
behave similarly to the galaxies we measured at z $\sim$ 1.2, galaxies would contribute less than 20\% as 
much ultraviolet light as would quasars, and the flux estimates from both sources would be consistent 
with the predictions of the proximity effect.  
	
\section{CONCLUSIONS}

Based on observations of eleven bright blue galaxies with 
redshifts 1.1 $<$ z $<$ 1.4, {\it no} Lyman continuum was detected.  
A stringent, model-independent, upper limit $<$ 2$ \times $10$^{-19}$ ergs cm$^{-2}$ sec$^{-1}$ 
$\AA^{-1}$ was obtained for the FUV flux escaping from eleven normal galaxies 
at redshifts of 1.1 $<$ z $<$ 1.4.
Given reasonable assumptions about the FUV flux from these galaxies, based on Bruzual-Charlot models, 
this represents less than 6\%, down to less than 1\%, of their intrinsic ionizing flux.  This 
corresponds to a relative escape fraction, f$_{rel~esc}$ $\le$ 1\%.
The magnitude of this decline confirms the basic premise of ``Lyman break'' searches for 
galaxies at intermediate redshifts.    
The integrated contribution of galaxies to the total UV radiation field at z $\sim$ 1.2 
is inferred to be less than 20\%, and may well be less than 2\%.  

\acknowledgments
Support for proposal 8561 was provided by NASA through a grant from the Space Telescope Science Institute, which 
is operated by the Association of Universities for Research in Astronomy, Inc., under NASA contract NAS 5-26555.
We thank Dr. Charles Steidel for helpful discussions and his generous sharing of unpublished 
photometry on several galaxies.  

\clearpage




\clearpage

\begin{figure}
\plotone{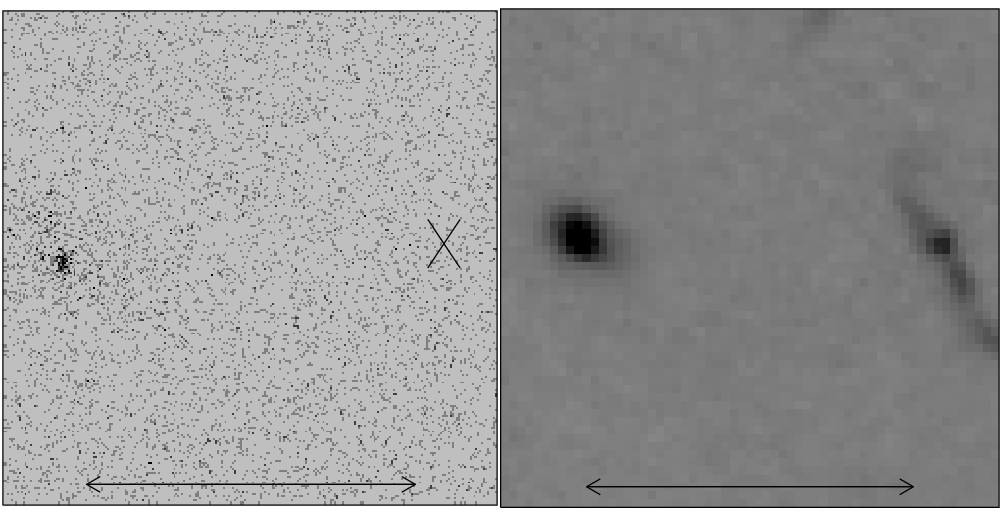}
\caption{Comparison of a WFPC2 visible image of SSA 22-16 and SSA 22-10 (right) 
with our STIS FUV exposure of the same field (left).  The images are adjusted 
to a common plate scale (the arrows are 4.5 arc seconds long) and are both oriented 
with north at top.  The lower-redshift galaxy SSA 22-10 (4.5" to the east of our target) is 
detected in the FUV but not our targeted galaxy SSA 22-16, which should lie within an 
arc second of the pointing center indicated by the ``X". \label{fig1}}
\end{figure}

\clearpage 

\begin{figure}
\plotone{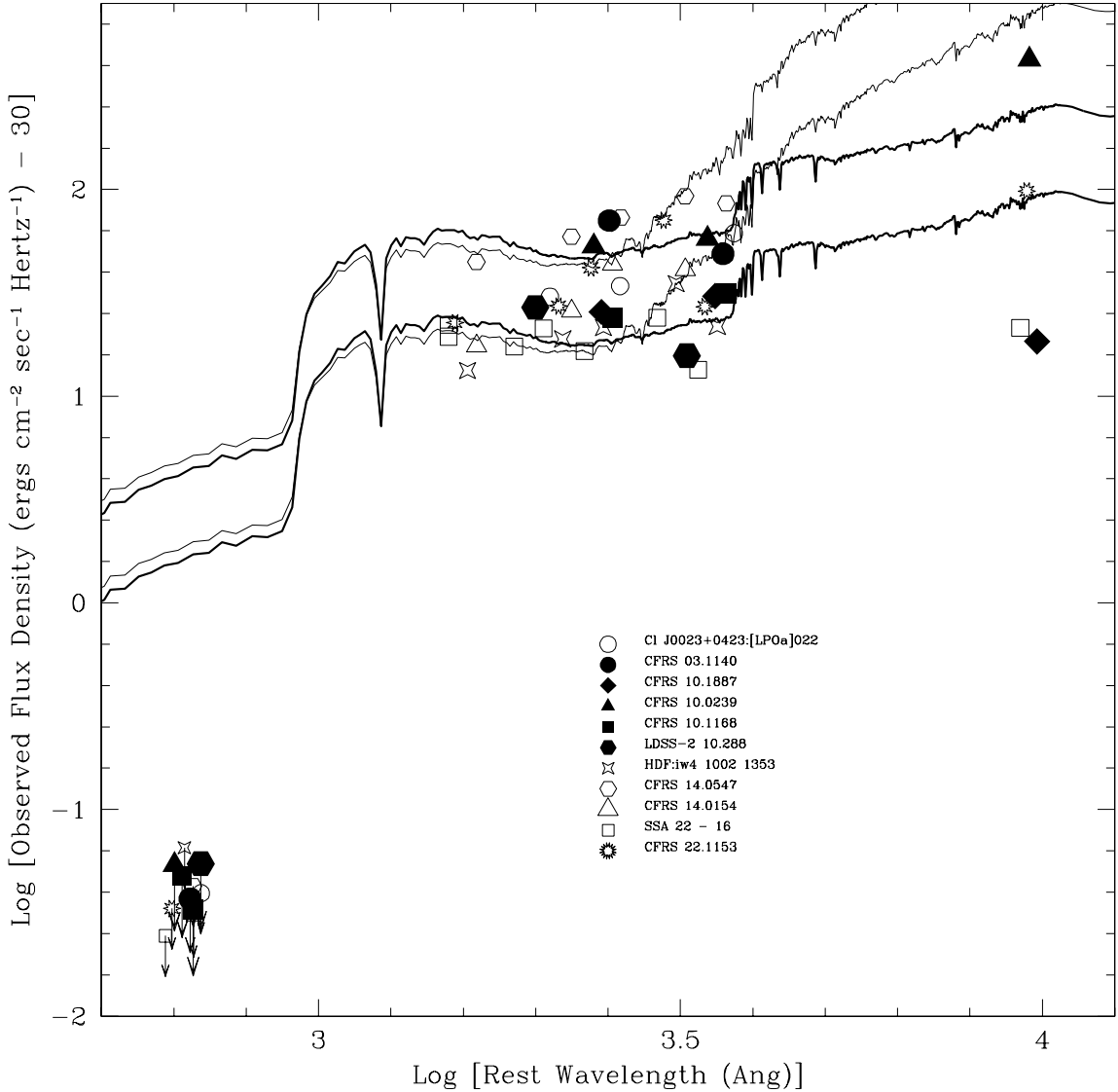}
\caption{Data points are dereddened 
spectral energy distributions (SEDs) for our 
target blue galaxies. Points at lower left, with down arrows, are new {\it upper limits} 
obtained in this paper.  
Optical and near-infrared photometry is taken from the literature.  Since error bars were usually 
not given for these data, we have adopted a standard uncertainty of
$\pm 10\%$, and scaled the vertical size of the plotted points to show this range.
For comparison, we have overplotted Bruzual-Charlot starburst models of evolving galaxies, 
redshifted to our rest frame, from an epoch of cosmic time corresponding to z $\sim$ 1.3. 
We assumed that the galaxies originally formed at z $\sim$ 10, and simulated 
both a strong and a weak secondary starburst starting at $z = 1.6$ 
(dark and thin lines, respectively).  These starbursts, assumed to have constant
star formation rate for one gigayear with a standard Salpeter IMF, produced 
the O and B stars observed in the UV.  Both models are plotted 
at two flux levels which bracket our data. \label{fig2}} 

\end{figure}

\clearpage

\begin{deluxetable}{crrrrr}
\tabletypesize{\scriptsize}
\tablecaption{Galaxies Targeted for Lyman Limit Imaging}
\tablewidth{0pt}
\tablehead{
\colhead{Galaxy} & \colhead{Redshift}   & \colhead{$\alpha$}   &
\colhead{Observation} &
\colhead{Exposure} &
\colhead{Galactic Bckgnd.}  \\
\colhead{} & \colhead{z} &
\colhead{$\delta$}     & \colhead{Date }  &
\colhead{Time}  & 
\colhead{@ 100$\mu$} \\
\colhead{} &  \colhead{} & \colhead{2000} & \colhead{(MJD) \tablenotemark{a}} 
& \colhead{(seconds)} & \colhead{(MegaJy/Sr)}  
}

\startdata
Cl J0023+0423: & 1.10740 & 00 23 51.00 & 51861.22 & 4000 & 2.5 \\

[LPO98a] 022   &         & 04 23 00.00 \tablenotemark{b} &     &    &  \\

CFRS 03.1140
& 1.18180
& 03 02 53.85
& 52199.04
& 6300
& 5.7  \\

&
& 00 13 04.00
&
&
&  \\

CFRS 10.1887 
& 1.2370
& 10 00 23.89
& 51970.04
& 5700 
& 2.8 \\
 
& 
& 25 18 12.00
&
&
& \\

CFRS 10.0239 
& 1.29190
& 10 00 40.24
& 51970.98
& 3700 
& 2.8 \\
 
&
& 25 11 05.00
&
& 
& \\

CFRS 10.1168
& 1.15920
& 10 00 42.09
& 51972.65
& 3500  
& 2.8 \\

&
& 25 14 41.00
&
&
&   \\

LDSS2 10.288 
& 1.108
& 10 46 32.85
& 52022.91
& 3900  
& 3.7 \\
 
&
& -00 11 41.60
&
&
& \\

HDF:iw4
& 1.221
& 12 36 27.24
& 51928.87
& 3700 
& 0.
\\

1002 1353
&
& 62 12 58.90
&
&  
& \\

CFRS 14.0547
& 1.160
& 14 18 03.47
& 52025.79
& 4100 
& 0.\\

&
& 52 30 21.70
&
& 
& \\

CFRS 14.0154
& 1.1583
& 14 18 17.80
& 51975.60
& 3700 
& 0.\\

&
& 52 26 31.00
&
& 
& \\

SSA 22-16
& 1.36
& 22 17 36.60
& 51834.49
& 5598 
& 4.7 \\

&
& +00 14 41.00
&
& (ACCUM)
&  \\

CFRS 22.1153 
& 1.31180
& 22 17 40.60
& 51828.66
& 3800 
& 4.7  \\
 
& 
& +00 18 21.80
&
&
& \\

\enddata

    
\tablenotetext{a}{Dates are given in Modified Julian Days (MJD, the Julian Date minus 2400000.5).}

\tablenotetext{b}{Position is known to within 5 arcseconds 
-- well within the field of view of the FUV MAMA detecrtors.}


\end{deluxetable}

\clearpage

\begin{deluxetable}{crrrrrrrrr}
\tabletypesize{\scriptsize}

\tablecaption{Limits and Photometry - Dereddening ($_0$) and Observed Fluxes 
\tablenotemark{a}}

\tablewidth{0pt}
\tablecolumns{10}

\tablehead{
\colhead{Galaxy} & 
\colhead{F$_{Lyco}$$<$}   & 
\colhead{U$_{n_o}$}   &
\colhead{B$_o$} &  
\colhead{G$_o$} &  
\colhead{V$_o$} &  
\colhead{R$_o$} &  
\colhead{R$_s{_o}$} &  
\colhead{I$_o$} &  
\colhead{K$_o$}  \\
\colhead{} & 
\colhead{F$_{Lyc}$$<$} & 
\colhead{U$_n$} &
\colhead{B}     & 
\colhead{G}     &  
\colhead{V}     &  
\colhead{R}     &  
\colhead{R$_s$} &  
\colhead{I} &  
\colhead{K}}

\startdata

Cl J0023+0423: \tablenotemark{b} & .0540 &     & 30.3 &     &  34.1 &     &     & 61.4   &   \\

[LPO98a] 022   & .0394      &     & 27.9 &     &  32.0       &    & &  59.2  &   \\

CFRS 03.1140  \tablenotemark{c}   & .0766  &     &      &     &  70.7 &     &     &  48.8  &   \\
	       & .0370      &     &      &     &  52.6 &     &     &  41.0  &   \\

CFRS 10.1887  \tablenotemark{d,h} 
& .0688 
& 
&
& 
& 24.5
& 
&
& 30.5
& 18.4
\\

& .0478
&
&
&
& 21.7
&
&
& 28.3
& 17.8
\\

CFRS 10.0239  \tablenotemark{d,h} 
& .0778
& 
&
& 
& 53.1
&
&
& 57.6
& 425.
\\

& .0541
&
&
&
& 47.1
&
&
& 53.5
& 411.
\\

CFRS 10.1168  \tablenotemark{d}
& .0470
& 
&
& 
& 23.0
&
&
& 31.4
&
\\

& .0327
&
&
&
& 20.4
&
&
& 29.1
&
\\

LDSS2 10.288  \tablenotemark{i}
& .0879
& 
& 26.9
&
&
& 15.7
& 
&
&
\\
& .0547
&
& 23.0
&
&
& 14.2
&
&
&
\\

HDF:iw4 \tablenotemark{f,j} 
& .0654
& 13.3
&
& 18.9
& 21.5
& 
& 35.0
& 21.7
&
\\

1002 1353
& .0654
& 13.3
&
& 18.9
& 21.5
& 
& 35.0
& 21.7
&
\\

CFRS 14.0547 \tablenotemark{e,g,j} 
& .0426
& 44.5
& 
& 59.2
& 73.3
& 
& 92.9
& 85.6
&
\\

& .0426
& 44.5
& 
& 59.2
& 73.3
& 
& 92.9
& 85.6
&
\\

CFRS 14.0154  \tablenotemark{e,j}  
& .0304
& 17.5 
&
& 25.8
& 53.6
& 
& 40.9
&
&
\\

& .0304
& 17.5 
&
& 25.8
& 53.6
& 
& 40.9
&
&
\\

SSA 22-16 \tablenotemark{j} 
& .0447
& 22.1
& 32.0
& 21.3
& 16.5
& 
& 24.8
& 13.4
& 21.4
\\

& .0245
& 15.6
& 24.5
& 16.3
& 13.5
&
& 21.4
& 11.9
& 21.0
\\

CFRS 22.1153  \tablenotemark{e,h,j} 
& .0606
& 22.7
&
& 27.3
& 41.8
& 
& 71.1
& 27.1
& 98.2
\\
& .0332
&  16.0
&
& 20.9
& 34.5
&
& 61.4
& 24.2
& 96.3
\\

\enddata

\tablenotetext{a}{$\times$ 10$^{-30}$ ergs cm$^{-2}$ sec$^{-1}$ Hz$^{-1}$}

\tablenotetext{b}{\citet{pos98}}

\tablenotetext{c}{\citet{ham95}}

\tablenotetext{d}{\citet{lef95}}

\tablenotetext{e}{\citet{lil95}}

\tablenotetext{f}{\citet{guz97}}

\tablenotetext{g}{\citet{bri98}}

\tablenotetext{h}{\citet{ham97}}

\tablenotetext{i}{\citet{gla95}}

\tablenotetext{j}{\citet{ste02}}

\tablecomments{For each object, photometry in the first row is dereddened (symbolized in the 
column headings by the subscript $_o$), while that in the second row is observed.  
U$_n$, G, and R$_s$ have central wavelengths 
of 3650, 4730, and 6830$\AA$, respectively.  B, V, I and K are standard Johnson filters.
Although error bars were not published, we assume that they are approximately 10\% for
most of these measurements.}

\end{deluxetable}

\end{document}